\documentclass[conference]{IEEEtran}
\IEEEoverridecommandlockouts

\usepackage{url}
\usepackage{hyperref}
\usepackage{graphicx}
\usepackage{subcaption}
\usepackage{todonotes}
\usepackage{listings}
\usepackage{markdown}

\begin{document}

\title{Software engineering to sustain a high-performance computing scientific application: QMCPACK}

\author{
\IEEEauthorblockN{William F. Godoy, Steven E. Hahn, Michael M. Walsh, Philip W. Fackler, Jaron T. Krogel, Peter W. Doak, \\ Paul R. C. Kent}
\IEEEauthorblockA{ 
Oak Ridge National Laboratory\thanks{This manuscript has been authored by UT-Battelle, LLC, under contract DE-AC05-00OR22725 with the US Department of Energy (DOE). The publisher acknowledges the US government license to provide public access under the DOE Public Access Plan (\url{https://energy.gov/downloads/doe-public-access-plan}).} \\
\{godoywf\}, \{hahnse\}, \{walshmm\},  \{facklerpw\}, \{krogeljt\},  \{doakpw\},  \{kentpr\}@ornl.gov} 

\IEEEauthorblockN{Alfredo A. Correa} 
\IEEEauthorblockA{Lawrence Livermore National Laboratory
\thanks{Performed under the auspices of the U.S. Department of Energy by Lawrence Livermore National Laboratory under Contract DE-AC52-07NA27344.}
}
correaa@llnl.gov

\IEEEauthorblockN{Ye Luo, Mark Dewing} 
\IEEEauthorblockA{Argonne National Laboratory}
\{yeluo\}, \{mdewing\} @anl.gov
}

\maketitle

\begin{abstract}
We provide an overview of the software engineering efforts and their impact in QMCPACK, a production-level ab-initio Quantum Monte Carlo open-source code targeting high-performance computing (HPC) systems. Aspects included are: (i) strategic expansion of continuous integration (CI) targeting CPUs, using GitHub Actions runners, and NVIDIA and AMD GPUs in pre-exascale systems, using self-hosted hardware; (ii) incremental reduction of memory leaks using sanitizers, (iii) incorporation of Docker containers for CI and reproducibility, and (iv) refactoring efforts to improve maintainability, testing coverage, and memory lifetime management. We quantify the value of these improvements by providing metrics to illustrate the shift towards a
predictive, rather than reactive, sustainable maintenance approach. Our goal, in documenting the impact of these efforts on QMCPACK, is to contribute to the body of knowledge on the importance of research software engineering (RSE) for the sustainability of community HPC codes and scientific discovery at scale.
\end{abstract}

\begin{IEEEkeywords}
Research software engineering, RSE, QMCPACK, CI, software sustainability, high-performance computing, HPC, sanitizers, GitHub Actions
\end{IEEEkeywords}


\section{Introduction}

Improving all aspects of scientific software development has become essential for the sustainability and trustworthiness of modern science~\cite{osti_1846008}. Early on, software
development for high-performance computing (HPC) was identified as a non-trivial task that requires a deep understanding of the
application and the targeted architectural systems~\cite{parashar1994study}. The complexity of HPC software is expected to increase
with the hardware heterogeneity characterizing the present systems~\cite{osti_1473756}. Therefore, understanding this
landscape~\cite{grannan2020understanding} is essential to guarantee the effective and strategic use of industry-standard software engineering
practices that have a positive impact on a particular HPC community target~\cite{basili2008understanding}. 

QMCPACK\footnote{\url{https://github.com/QMCPACK/qmcpack}}~\cite{kim2018qmcpack,doi:10.1063/5.0004860} is an open-source many-body
ab-initio Quantum Monte Carlo~\cite{foulkes2001} (QMC) framework. QMC methods solve directly the Schrodinger equation offering more accuracy than density functional theory (DFT) methods at a greater computational cost. QMCPACK is used to compute the electronic structure of atoms, molecules, 2D nanomaterials and solids.
The code supports many architectures ranging from laptops through mid-range clusters to the largest supercomputers in the world. Written in C\texttt{++} with a modular design, it uses a hybrid parallel approach combining the message passing interface (MPI) with multicore, many-core CPUs and GPUs. It primarily uses OpenMP~\cite{chapman2008using} for CPU and GPU execution, with additional CUDA~\cite{nickolls2008scalable},
HIP/ROCm~\cite{hip} and SYCL~\cite{reyes2016sycl} for higher GPU performance. Releases are typically made quarterly. QMCPACK is part of the application development portfolio~\cite{osti_1594850} of the US Department of Energy Exascale Computing Project (ECP)~\cite{8528398}. ECP's mission 
addresses currently intractable problems of strategic importance and national interest at the same time as the first exascale systems, such as Frontier\footnote{\url{https://www.olcf.ornl.gov/frontier}}, are deployed.

Cloud technologies have contributed greatly to the development and empowerment of the open-source community in the last decade. They provide a rich ecosystem and resources that facilitate the application of modern software engineering practices to enable the trustworthiness that is crucial in scientific code~\cite{fox2011cloud}. Code hosting services like GitHub have evolved into a one-stop shop that adds value to the modern software development process. One example is the rapid adoption of GitHub Actions~\cite{decan2022use} since its initial release in 2019, providing computational resources for continuous
integration/continuous development (CI/CD) workflows. It also provides seamless integration with third-party cloud services
(\textit{e.g.}, code analysis, coverage, reporting) via a marketplace of reusable workflow ``actions''.

In this paper, we describe our efforts and the impact of incorporating modern software engineering practices in the QMCPACK code. The goal of these changes has been to
improve the quality of the software and enable significant refactoring while also keeping the barrier for open source software contributions low.
Metrics are provided to quantify the value of such practices in a real HPC application with active users and developers. We showcase: (i) the evolution
and failure rates on our GitHub Actions-based CI system to cover CPU and GPU test cases, (ii) the code sanitization process to
resolve and future-proof for memory leaks, (iii) the inclusion of Docker containers to enable reproducible virtual testing
environments, (iv) code refactoring efforts to ensure the sustainability of the scientific efforts around the software. As a result of these improvements, the QMCPACK community benefits from a robust first line of quality checks ahead of deployment, testing, and science production on HPC systems.

The paper is organized as follows: Section~\ref{sec:Background} provides background information on the QMCPACK software.
Section~\ref{sec:Software-Improvements} describes the software improvements showcased in this paper along with metrics for each
covered aspect. Related efforts in applying software engineering practices to HPC codes are presented in Section~\ref{sec:Related-Work}. Section~\ref{sec:Conclusions} presents the conclusions of this work, while an artifact description
for the activities and reproducibility of the metrics reported in this paper is given in Appendix~\ref{ap1:Artifact}.


\section{Background}
\label{sec:Background}

QMCPACK is an open-source Quantum Monte Carlo framework with a sizeable number of users as evidenced by the number of
citations of the reference paper by Kim et al.~\cite{kim2018qmcpack}, 223 as of April 2023 according to Google Scholar.
Development
started as early as 1998 and has involved many stages of development and modernization through the years from different teams of developers~\cite{doi:10.1063/5.0004860}. QMCPACK's current development process ensures that the code performs on several heterogeneous architectures targeting a wide range of parallel computing systems. The latest efforts include its readiness on novel Arm-powered test bed clusters~\cite{10.1145/3581576.3581621}.

Table~\ref{tab:software} presents some of the characteristics of the QMCPACK software infrastructure:

\begin{table}[!ht]
\centering
\caption{Software characteristics of QMCPACK}
\begin{tabular}{| p{0.35\linewidth} | p{0.55\linewidth}|}
\hline
Software characteristic  & Value \\
\hline
Repository           & \url{https://github.com/QMCPACK/qmcpack} \\ 
Open-source License  & UIUC / NCSA \\
Languages            & C\texttt{++}17, C, Python 3 \\
Parallel models      & CUDA, HIP, OpenMP, SYCL, MPI \\
Build system         & CMake $\geq$ 3.17 \\ 
Lines of code        &  $\sim 2\times10^5$       \\
Tests suites         & unit, deterministic, stochastic, performance \\
Dependencies         & Libxml2, Boost, HDF5, BLAS, LAPACK, FFTW3, MPI \\ 
Current version      & 3.16.0 \\
\hline
\end{tabular}
\label{tab:software}
\end{table}

Compilation of the QMCPACK C\texttt{++} source code generates a binary executable per selected configuration, \texttt{qmcpack} (real number calculations) and
\texttt{qmcpack\_complex} (complex number calculations). Users can configure and build the code via CMake, but the preferred installation route is through the 
Spack~\cite{gamblin2015spack} package manager designed specifically for supercomputers and easy installation for multiple configurations on multiple operating systems. QMCPACK ships with the Nexus workflow management system~\cite{KROGEL2016154} to ease the running of research workflows in the
creation of inputs and data analysis of outputs. Nexus acts as a front end to QMCPACK, both by abstracting the interface to the application via Python calls and facilitating running QMCPACK at scale on production HPC systems.

QMCPACK's current software process can be summarized as follows: 

\begin{itemize}
    \item GitHub ``fork and pull request'' contribution workflow.
    \item Continuous integration (CI) checks on many build combinations are enforced.
    \item Code contributions must be reviewed by the core team, CI checks passed, and the final merge must be done by someone at an institution not involved with the contribution.
    \item Open source compiler and library dependencies are supported for versions released in the last two years.
    \item Security: two-factor authentication for all members and manually triggered self-hosted CI by a few admins.
    \item User documentation hosted on \texttt{readthedocs}\footnote{\url{https://qmcpack.readthedocs.io/}}.
    \item Tests that take too long to run practically in CI are run nightly on several systems and results reported on cdash\footnote{\url{https://cdash.qmcpack.org/}}.
    \item Communication is done via project Slack, GitHub issues, and regular project meetings.
\end{itemize}

QMCPACK is heavily developed and maintained by National Laboratory staff members, as indicated by the authors' affiliations. Much of the work is carried out by the core team of domain scientists in QMC methods, in collaboration with computer scientists in HPC, and those identified as RSEs at National Laboratories~\cite{10071971}. While there is no formal RSE career track, the addition of staff dedicated to software engineering tasks from Oak Ridge National Laboratory~\cite{10078175} has been welcomed in executing the tasks described in the present paper towards meeting QMCPACK's sustainability goals.


\section{Software Improvements}
\label{sec:Software-Improvements}

This section describes the software engineering efforts of the last two years. Our goal is to improve and sustain the scientific community efforts around QMCPACK's evolution in the current and future landscape of HPC.

\subsection{Docker containers}

Containers are a lightweight alternative to virtual machines to create portable and reproducible environments without a full operating system.
Due to its nature as a complex physical simulation framework, QMCPACK has several configuration options to build the final executable.
We use pre-packaged Docker images to provide uniformity across environments used in GitHub Actions CI tests, including appropriate dependencies hosted in DockerHub.
In addition, ``Dockerfiles'' used for image configuration are available in the QMCPACK repository for reproducibility.
These containers can be pulled to any Linux host to trigger an interactive debug session in the same environment used for CI.
On the other hand, we do not use containers for macOS and GPU runs as there is currently limited value due to the small number of targeted use cases and the bare-metal nature of testing on GPUs. 

In addition, we have begun exploring using the Spack package manager to generate Docker images and leverage its package managing capabilities.
This integration and interoperability simplify the distinct processes each dependency may require to build and install correctly on different operating systems and environments.

\subsection{Strategic CI expansion on GitHub Actions}

Starting in mid-2021, QMCPACK began the adoption of GitHub Actions in place of Jenkins for CI.
The main reasons are:
(i) making full utilization of the freely provided GitHub CPU runners in virtual machines (VMs) using our Docker containers, and
(ii) providing CI for GPU development via runners hosted at ORNL equipped with NVIDIA and AMD cards. In addition, GitHub Actions offers simple workflow configurations via YAML files, provides reusable automated functionality via open-source workflow ``actions'' and other services (e.g., Codecov, DockerHub, GitHub Actions data encryption via ``secrets''), and enables data mining access of the logs via the GitHub command line interface (CLI). This adoption allowed for the desired sustainability goals of lowering maintenance costs while adding relevant validation and verification checks to QMCPACK.

QMCPACK uses GitHub Actions VMs to perform an automatic initial check for CI using Docker containers on available virtual machines (VMs) for the configurations listed in Table~\ref{tab:GitHubActions}.
While the main target is Linux, macOS testing using the Accelerate framework for linear algebra is also provided.
The current set of CI configurations has been evolving to prioritize checks that address problematic and error-prone areas of the code.
As such, address (ASan) and undefined behavior (UBSan) sanitizer checks and code coverage using GNU's gcov were added early in the CI migration, while jobs running in serial mode (NoOMP, NoMPI) were added later. 
The latter ensures the code can be built with the fewest number of dependencies and in the easiest mode to utilize debuggers and profilers.
An important benefit is that offloading the hardware maintenance and system administration aspects lowers the overall cost of adding more configurations.
The trade-off is that while each job runs on limited hardware resources at small scales, they are all triggered concurrently when new contributions are opened via pull requests targeting the main development branch. The configurations have been curated to keep turnaround in the CI at close to one hour of real-time.
 
\begin{table}[!ht]
\centering
\caption{CI on GitHub Actions hosted systems (CPU-only) as of March 2023}
\begin{tabular}{||p{0.35\linewidth} | p{0.55\linewidth}||}
\hline
System VM     & Job Configuration  \\
\hline \hline 
\textit{linux}          & GCC9-NoMPI-Debug-Real         \\
2-core CPU (x86\_64)    & GCC9-NoMPI-NoOMP-Real           \\
7GB RAM, 14GB SSD       & GCC9-NoMPI-NoOMP-Complex        \\
                        & GCC9-NoMPI-Sandbox-Real         \\
                        & GCC9-MPI-Gcov-Real              \\
                        & GCC9-MPI-Gcov-Complex           \\
                        & GCC11-NoMPI-Werror-Real       \\
                        & GCC11-NoMPI-Werror-Complex       \\
                        & GCC11-NoMPI-Werror-Real-Mixed       \\
                        & GCC11-NoMPI-Werror-Complex-Mixed       \\
                        & Clang10-NoMPI-ASan-Real       \\
                        & Clang10-NoMPI-ASan-Complex       \\
                        & Clang10-NoMPI-UBSan-Real      \\
                        & Clang12-NoMPI-Offload-Real       \\
\hline
\textit{macOS}          & GCC11-NoMPI-Real    \\
3-core CPU (x86\_64)    & \\
14GB RAM, 14GB SSD      & \\
\hline
\end{tabular}
\label{tab:GitHubActions}
\end{table}

Testing each pull request requires running the proposed changes on GPUs.
This is achieved via a second-level CI workflow, which must be triggered manually (by authorized members).
The specifications for each system are provided in Table~\ref{tab:self-hosted} along with the job configurations run on each system (sulfur and nitrogen). At the time of this work the current systems had a Red Hat Enterprise Linux (RHEL) operating system version 8.
This strategy meets our security requirements to prevent access to unauthorized agents, thus adding another layer of security on top of GitHub's two-factor authentication and branch
protections.
It is worth noting that the CPU and GPU testing runs concurrently.
Due to limited GPU memory, GPU jobs are, however, run sequentially.
To provide a partial test of the OpenMP target offload GPU implementation  in the first tier of CI we utilize the ability of OpenMP offload code to be compiled by LLVM for host CPUs.
This provides an initial screen for modifications to the GPU code without needing to commit any GPU hardware.

\begin{table}[!h]
\centering
\caption{CI on ORNL self-hosted systems as of March 2023}
\begin{tabular}{||p{0.29\linewidth} | p{0.58\linewidth}||}
\hline
System Specs      & Job Configuration  \\
\hline \hline 
\textit{sulfur Linux RHEL8}     &                                           \\
                    &                                           \\
\textbf{CPU}        & GCC8-NoMPI-MKL-Real-Mixed                 \\
2xIntel Xeon        & GCC8-NoMPI-MKL-Complex-Mixed              \\
Gold 6248R 24-core  & GCC8-NoMPI-MKL-Real                       \\
                    & GCC8-NoMPI-MKL-Complex                    \\
                    &                                           \\
\textbf{GPU}        &                                           \\
NVIDIA Tesla V100   & Clang15-MPI-CUDA-AFQMC-Offload-Real-Mixed \\
                    & Clang15-MPI-CUDA-AFQMC-Offload-Real       \\
                    & Clang15-MPI-CUDA-AFQMC-Offload-Complex-Mixed  \\
                    & Clang15-MPI-CUDA-AFQMC-Offload-Complex    \\
                    & Intel19-MPI-CUDA-AFQMC-Real-Mixed         \\
                    & Intel19-MPI-CUDA-AFQMC-Complex-Mixed      \\
                    & Intel19-MPI-CUDA-AFQMC-Real               \\
\hline
\textit{nitrogen Linux RHEL8}  &                  \\
2xAMD EPYC          & \\
7542 32-Core        & \\
                    &                                \\
\textbf{GPU}        &                                \\
NVIDIA Tesla V100   & GCC8-MPI-CUDA-AFQMC-Real-Mixed \\
                    & GCC8-MPI-CUDA-AFQMC-Real  \\
                    & GCC8-MPI-CUDA-AFQMC-Complex-Mixed \\
                    & GCC8-MPI-CUDA-AFQMC-Complex  \\
                    &                                           \\
\textbf{GPU}        & \\ 
AMD Vega 20         & ROCm-Clang13-NoMPI-CUDA2HIP-Real-Mixed \\
                    & ROCm-Clang13-NoMPI-CUDA2HIP-Real \\
                    & ROCm-Clang13-NoMPI-CUDA2HIP-Complex-Mixed \\
                    & ROCm-Clang13-NoMPI-CUDA2HIP-Complex \\
\hline
\end{tabular}
\label{tab:self-hosted}
\end{table}

To quantify the impact of the preventive nature of the CI system, we measure the total number of occurrences and the failure rates per month since the beginning of this effort.

\begin{figure}[ht]
    \centering
    \subfloat[GitHub Actions CI.]{
        \includegraphics[width=3.55in]{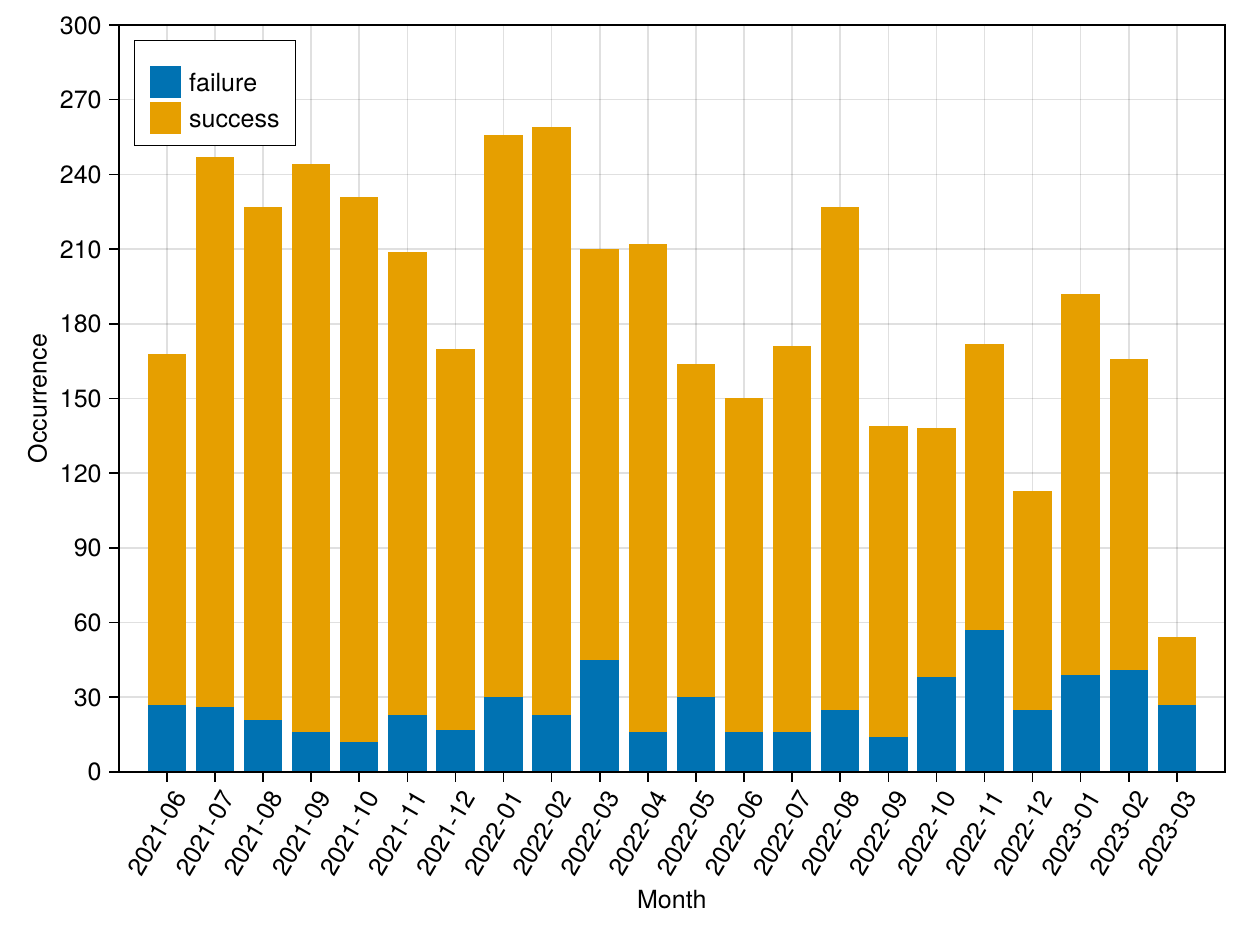}
        \label{fig:totals_gha}
    }
    \vspace{0.1in}
    \subfloat[Self-hosted CI.]{
        \includegraphics[width=3.55in]{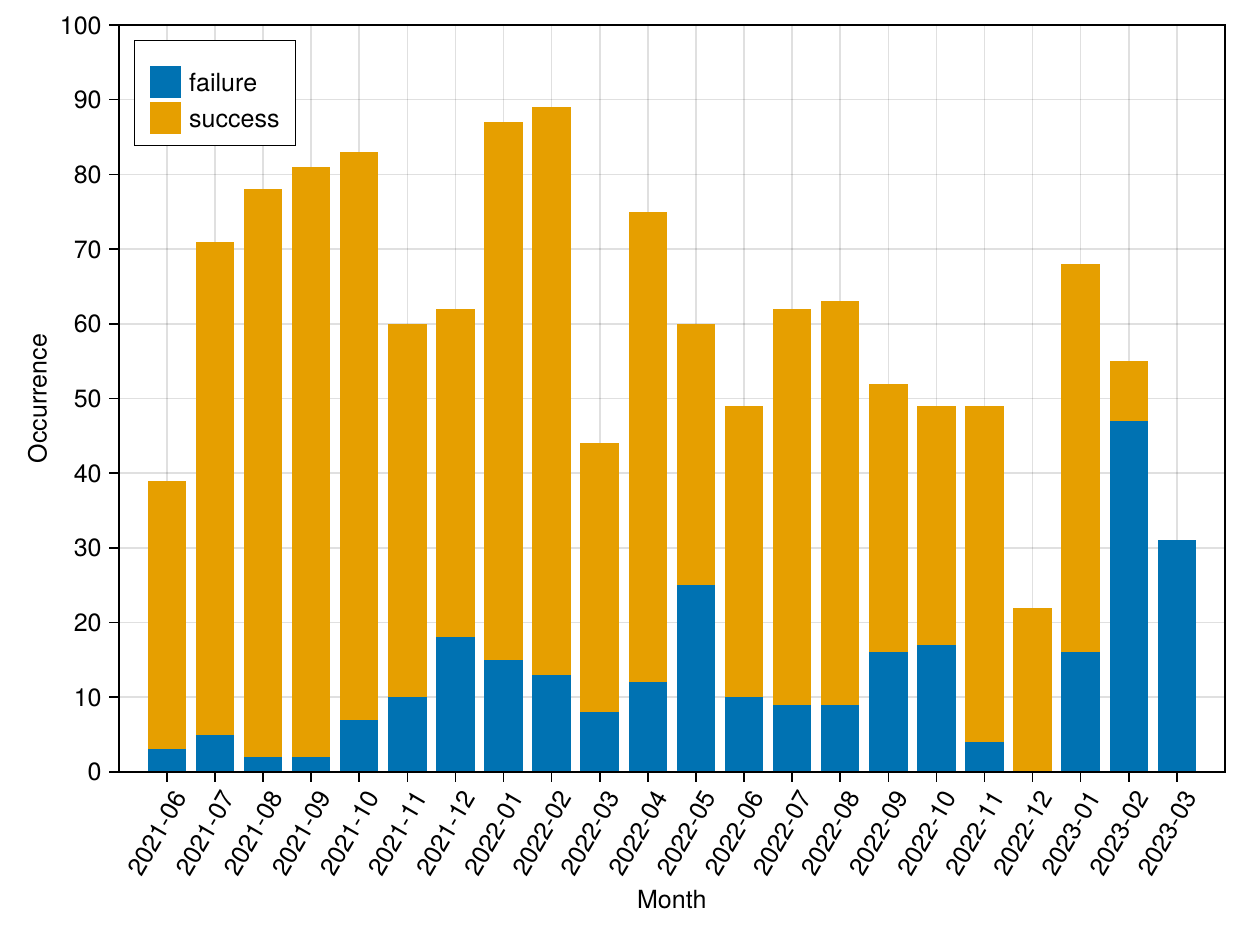}
        \label{fig:totals_ornl}
    }
    \caption{Total workflows.}
    \label{fig:totals}
\end{figure}

Figure~\ref{fig:totals} shows the total number of workflows (a collection of jobs) run in each month, also as an indication of the volume of development activity.
Note the difference in scale (100 \textit{vs.} 300) between the number of workflows run on self-hosted ORNL runners (Fig. \ref{fig:totals_ornl} ``ornl CI'') and those run on the free GitHub-hosted runners (Fig.
\ref{fig:totals_gha} ``GitHub Actions CI'').
This is explained due to our loose requirement for passing GitHub Actions CI first-level checks prior to using the self-hosted runners, which are a more constrained resource.

Failure rates can be extremely variable, and in the case of cloud-based CI in Figure~\ref{fig:totals_gha}, it may also be dependent on external factors (\textit{e.g.}, Codecov failed requests or GitHub Actions incidents). Figure~\ref{fig:totals_ornl} illustrates the importance of having GPU CI, especially around times of high development activity. 

Testing implicitly touches the entire software stack (the Nexus workflow system, various wavefunction conversion utilities, etc), not only in the main QMCPACK application.
The high failure rate experienced in the first quarter of 2023 reflects problems and updates in the test infrastructure.
Peaking in March of 2023, the failures are explained due to the process of upgrading both CI systems without pausing development while issues were resolved.
We introduced Spack-based CentOS CI on GitHub Actions runners, one configuration for which led to the discovery of a persistent bug that was determined to be caused by the OpenBLAS threading policy until a workaround was implemented in QMCPACK and promptly fixed by the OpenBLAS developers.
At the same time, our self-hosted runners had an operating system upgrade (from RHEL8 to RHEL9), which led to several refactoring efforts in sorting out compatibility of CPU and GPU dependency versions. The latter is a good case for how scientific software priorities might differ from other types of development.

\begin{figure}[ht]
    \centering
    \subfloat[``GitHub Actions CI'']{
        \includegraphics[width=3.55in]{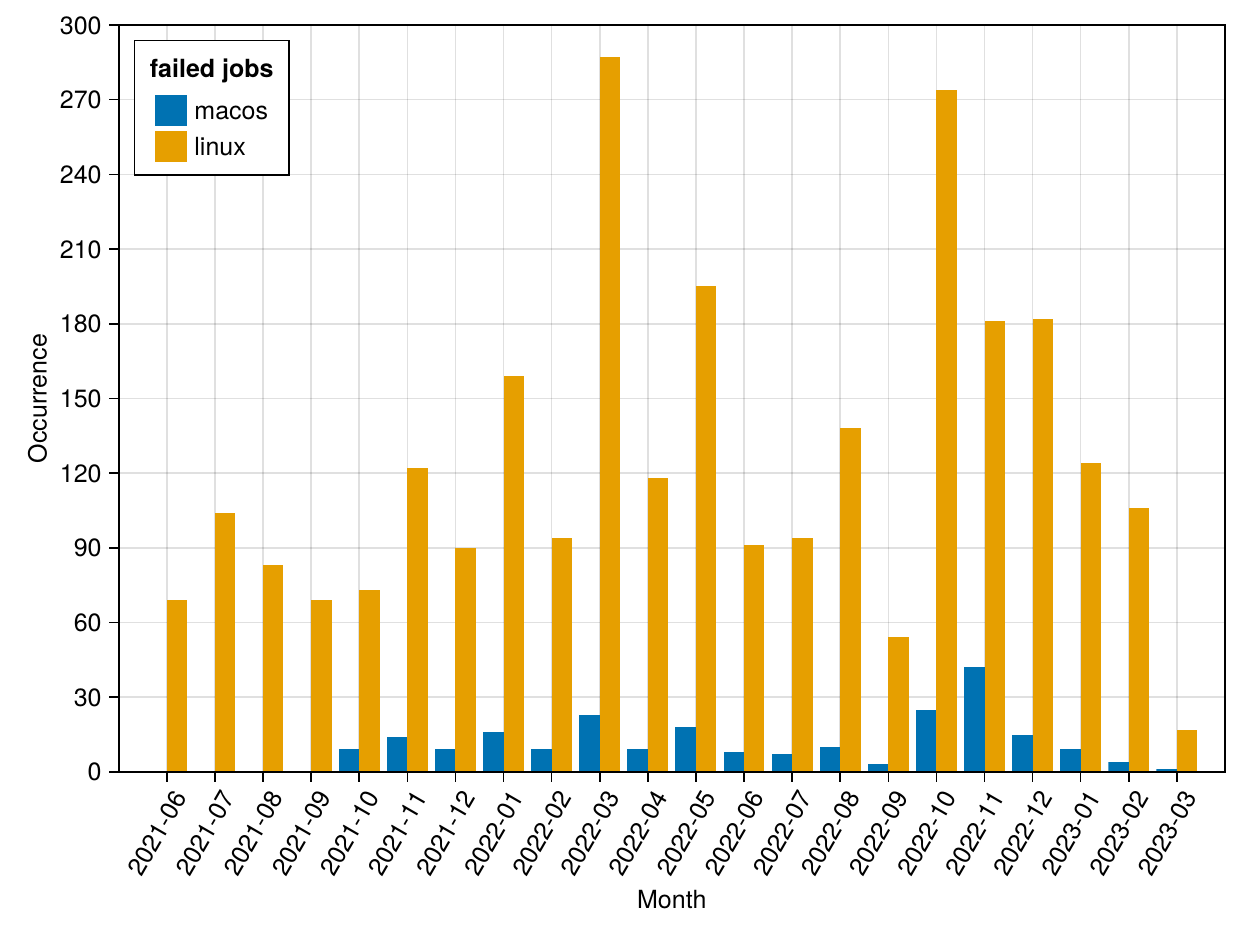}
        \label{fig:failed_jobs_gha}
    }
    \vspace{0.1in}
    \subfloat[``Self-hosted CI'']{
        \includegraphics[width=3.55in]{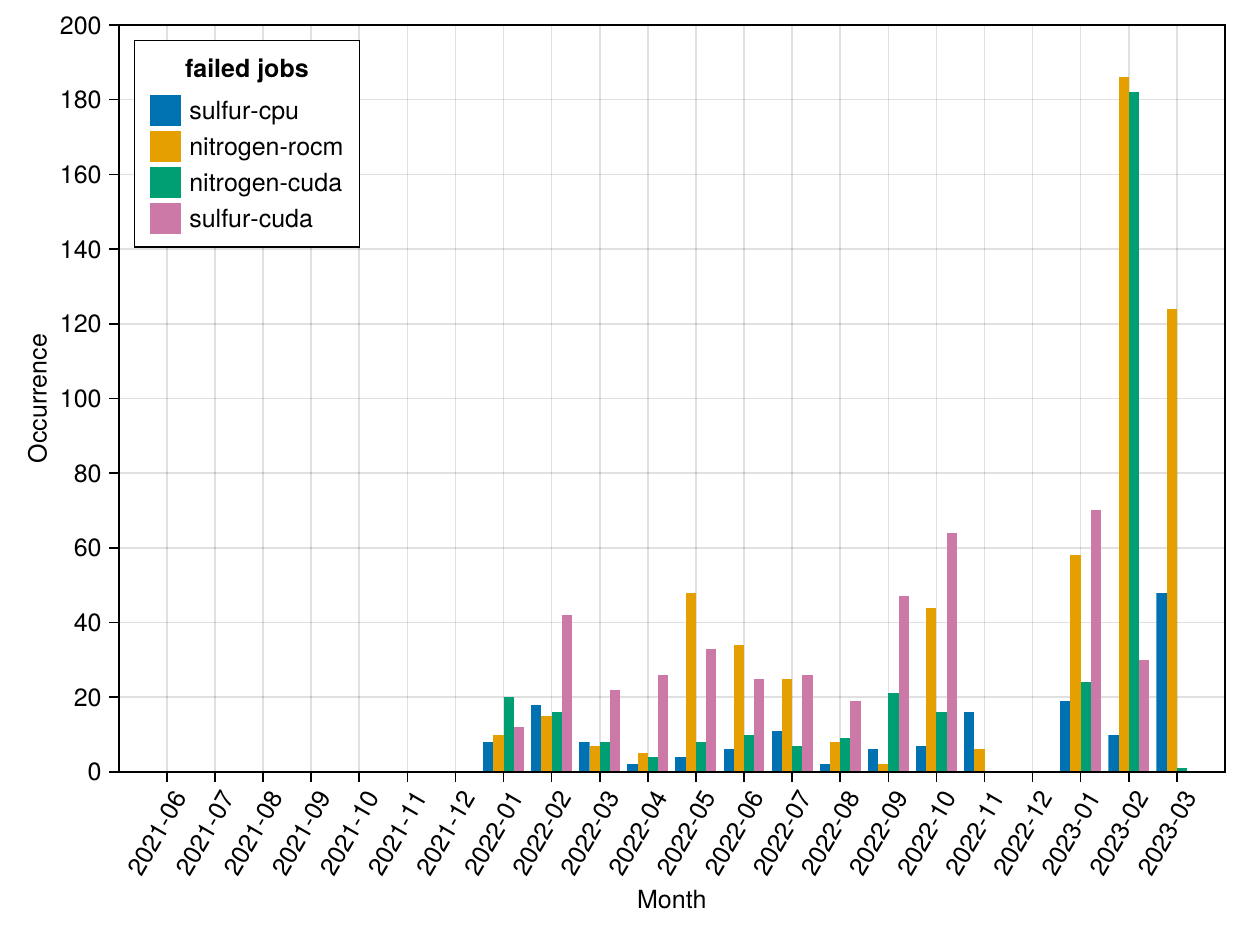}
        \label{fig:failed_jobs_ornl}
    }
    \caption{Failed jobs.}
    \label{fig:failed_jobs}
\end{figure}

To provide information on the nature of the failed CI jobs, Figure~\ref{fig:failed_jobs} shows the total number of failed jobs (not workflows) per month.
In the case for GitHub Actions CI, Figure~\ref{fig:failed_jobs_gha}, the majority of failed jobs come from configurations targeting the Linux runners.
Still, failures caught on the macOS free runners on GitHub Actions are not negligible and provide coverage where many new users initially experiment with the code.
It can be observed in Fig.~\ref{fig:failed_jobs_ornl} that since the beginning of 2022, the jobs identifiers were refactored to meet current standard names, similar to those on the longer-running nightly tests, so they can be classified by job name pointing at a particular configuration type as those shown in Table~\ref{tab:self-hosted}.
Failure counts illustrate the level of development activity on new features, high activity around NVIDIA GPU implementations (sulfur and nitrogen -cuda) early in 2022, while the second half indicates high levels of activity in AMD GPU implementation features (nitrogen-rocm).
GPU testing was able to catch several problems before they made it into the mainline development branch. 
Thus, these checks improve the reliability and robustness of QMCPACK while lowering overall development costs as actively developed GPU technologies targeting exascale systems continue to mature.

\subsection{Sanitizing critical code paths}

Memory safety has been a long-standing part of technical debt in legacy C and C\texttt{++} code~\cite{10.1145/3129743.3129745}.
Memory leaks, the lack of release of out-of-scope acquired resources, add pressure in the target operating system and threaten long-running simulations.
Through the compile-time instrumentation of address sanitizers~\cite{serebryany2012addresssanitizer}, several memory leaks were identified in QMCPACK unit and functional tests due to the use of raw pointers in pre-C\texttt{++}11 development.
To our knowledge, no memory safety checks were applied to QMCPACK in its early development stages.
We replaced raw pointer instances with ``smart pointers'' introduced in the C\texttt{++}11 standard library, \texttt{std::unique\_ptr} and \texttt{std::shared\_ptr}, to manage object lifetimes automatically following the \emph{resource allocation is initialization} - RAII - paradigm.
The overall process was incremental in nature as each leak provided an opportunity to refactor the affected
object lifetime to favor the use of ``unique'' pointer instances rather than the more permissive ``shared'' pointers.
As each memory leak was resolved, we reinstated each failing test in the GitHub Actions CI (ASan) jobs listed in Table~\ref{tab:GitHubActions} to future-proof critical code execution paths.
We favor the use of sanitizers due to smaller overheads compared to purely run-time approaches such as Valgrind memcheck. 

\begin{table}[h]
\centering
\begin{tabular}{|l|c|}
\hline
Metric type & Occurrences  \\
\hline
leaky unit tests                  & 10   \\
leaky functional tests            & 10   \\
direct leaks                      & 25   \\
indirect leaks                    &  1    \\
\hline 
\end{tabular}
\caption{Metrics showing the scope and nature of the resolved memory leaks in QMCPACK}
\label{tab:leaks}
\end{table}

Table~\ref{tab:leaks} shows the number of failures in the unit and deterministic tests run on CI.
There are two types of leaks to be addressed:
(i) direct, in which allocation and deallocation must be done in QMCPACK, and 
(ii) indirect, in which allocation and deallocation request is done outside QMCPACK, but triggered through the use of public functions in a dependency, e.g. libxml2.
Further details can be found by tracking the issue link in Appendix~\ref{ap1:Artifact}.
The vast majority of these leaks are of a ``direct'' nature.
There were no cases where we had to fall back to shared ownership.
Due to the existence of interoperable C\texttt{++} and C code in QMCPACK, we supplied user-defined custom destructors to \texttt{std::unique\_ptr} to properly free C-style structs.
The only indirect leaks that were resolved were related to the use of C functions in the \texttt{libxml2} library dependency in which the lack of destructor calls was addressed.

\subsection{Code refactoring}

We list some of the important aspects that are currently targeted in QMCPACK's strategic code refactoring efforts that will result in a greater return for the community.

\paragraph{Checkpoint-restart} software running at scale should be able to recover from failure or continue running beyond a maximum allowable time.
QMCPACK uses a checkpoint-restart approach where the state of the simulation is regularly stored to disk. Checkpoint files contain enough data written as plain text, XML and HDF5 files to restart mid-run and are deterministic so as not to affect the results.
Since QMCPACK uses Monte Carlo, this includes restarting of the random number generators.
Part of the effort also includes refactoring I/O components in terms of object ownership and lifetime and improved reusability in the interactions with the underlying HDF5 library.

\paragraph{Phasing out legacy GPU functionality.}
As part of the ECP application development effort, QMCPACK developed a new design and strategy to exploit multiple architectures in a performance-portable manner as described by Luo et al.~\cite{10024599}.
During the development stage of the new performant drivers, users could rely on the original CUDA-specific GPU implementation.
This ``legacy'' implementation was incompatible with CPU-only drivers and resulted in entirely separate calls in the wavefunction calculation phase.
Once the new drivers reached an acceptable level of performance and maturity, the legacy drivers were completely removed, with a reduction of nearly $40\,K$ lines of code.
This illustrates the importance of a software evolution strategy that is able to meet the demands of the science that comes out of new HPC available hardware.
In QMCPACK's case, this strategy is expected to pay off as more exascale and heterogeneous systems continue to be deployed.
By maximizing the number of lines of common code between architectures, maintenance costs are reduced and higher code quality can be expected.

\begin{figure}[ht]
    \centering
    \includegraphics[width=3.55in]{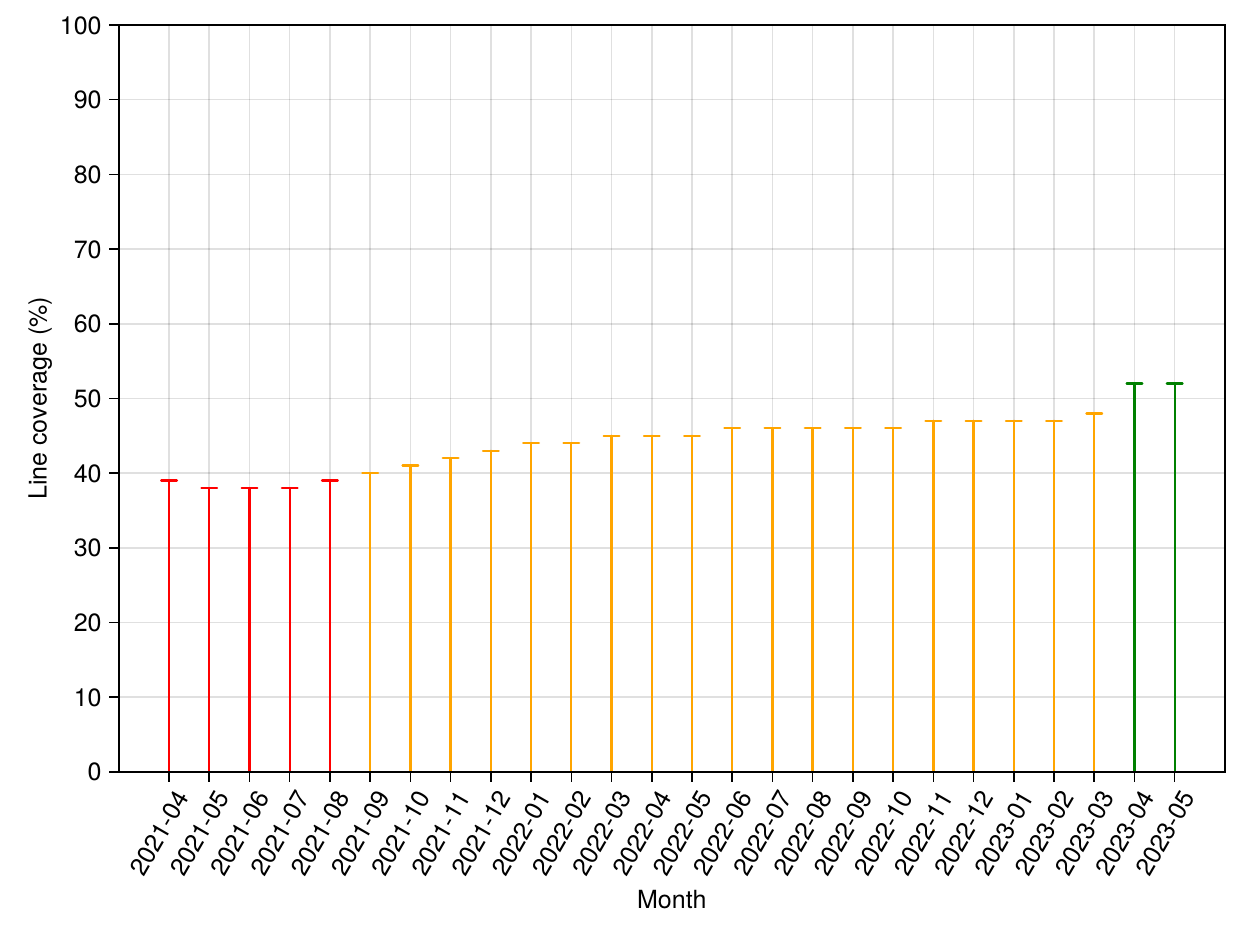}
    \caption{Monthly evolution of QMCPACK code line coverage reported by CI gcov tests.}
    \label{fig:coverage}
\end{figure}

\paragraph{Code coverage via unit tests}
QMCPACK's test suite on GitHub Actions CI uses gcov builds and reports the results to
Codecov to track the progress for code coverage in our tests. 
Pull request contributors will be notified how the new code affects current coverage levels. Decreasing coverage levels will issue a failed CI status.
Code coverage has been increasing since the addition of these CI checks from 38\% (absolute) to 52\% almost linearly with time as shown in Figure~\ref{fig:coverage}.
This coverage (complete line coverage) is expected to keep increasing and already includes all the core functionality of the application. It has been sufficient to catch problems in the greater software stack, although it does not include GPU offload cases. This metric is sensitive to the gcov version, as shown in Figure~\ref{fig:coverage}, the increase in the last two months is due to a GCC toolchain upgrade from version 9 to 11.
While not a specific goal but rather a good practice, progressive migration of code coverage from functional to unit tests is part of our long term goals.
We aim to target the scope of functional tests towards correctness and science cases;
unit tests cover the intended basic software functionalities.

\paragraph{Input validation} QMCPACK's native input is structured in XML files.  While this choice allows for a close relationship between the input data and the underlying modular class framework, the distributed parsing of input on a class by class basis complicates standardization of parsing and validatation of input data.  In concert with the development of performance portable drivers, an improved means of standardized input parsing has been introduced.  In the improved implementation, a fully typed input specification is encoded in association with each underlying class and parsing is deferred to a common base class.   This design facilitates transparent and more complete validation of input data with resulting benefits of application robustness to end users. 

\paragraph{Compile-time to runtime variants} as seen in Tables~\ref{tab:GitHubActions} and ~\ref{tab:self-hosted}, QMCPACK has a large number of compile-time configuration options that narrows the applicability of the resulting binary to specific cases.
These were introduced early in the development of the application.
With the availability of rich meta-programming capabilities via C\texttt{++} templates, we aim to migrate
some of these compile-time configurations to runtime variants.
Thus, a single executable can support several of these configurations without the need to recompile or distribute several versions of the code.
The most critical case is the unification of the use of complex and real numbers in wavefunction calculations.
This is an ongoing effort that would simplify use of the application, reduce testing burden, and simplify installation/distribution.


\section{Related Work}
\label{sec:Related-Work}

This section collects reported works on the intersection of HPC and software engineering which is relevant to the role and the increased complexity of RSE activities.
As described recently by Grannan et al.~\cite{grannan2020understanding} HPC software projects have evolved to greatly take advantage of software engineering practices.
McInnes et al.~\cite{mcinnes2021community} propose embracing community software ecosystems so HPC projects can achieve their science goals through enriching the synergies with other communities with similar software needs.
In this regard, Bartlett et al.~\cite{10.14529/jsfi170104}, and Heroux et al.~\cite{heroux2020e4s} present the xSDK and E4S ecosystems, respectively, that provide community guidelines and policies towards a coordinated software interoperability, performance portability, and sustainability strategy.

Carver et al.~\cite{4222616} provide lessons learned from surveying five HPC applications and the gaps and opportunities for applying software engineering practices.
One important lesson is the difficulty of verification and validation.
Schmidberger and Brügge~\cite{6341492} conducted a survey among HPC projects that highlighted the trade-offs between software engineering practices and the time and resources required as one reason preventing further adoption.
Dubey et al.~\cite{dubey2014evolution} describes the symbiotic relationship between scientific research and good software engineering on the widely used FLASH code.
Lessons learned on the development of GPU applications are presented by van Werkhoven et al.~\cite{van2020lessons}.
In their work, they argue that GPU development comes with specific challenges and that code must be of sufficient quality.
In a recent paper, Pachev et al.~\cite{10.1145/3491418.3535124} present a CI framework for HPC applications.
Their work is based on GitHub Actions which cites similar advantages for the use of self-hosting CI runners as in this work.

At the beginning of the last decade, Schmidberger and Schmidberger~\cite{6341491} proposed the concept of ``software engineering as a service'' for HPC to lower existing barriers. Heroux et al.
~\cite{10.1007/978-3-030-44728-1_6} propose an iterative workflow, Productivity and Sustainability Improvement Planning (PSIP), that allows HPC software teams to identify development bottlenecks. Eisty et al.~\cite{10.1007/978-3-030-50436-6_33} used this methodology to develop a testing framework to test non-deterministic parallel research software.
Antonioletti et al.~\cite{antonioletti2000software}, anticipated the rapid evolution of software engineering practices for HPC codes as we entered the new century. More recently, Heroux~\cite{10078171} proposes the concept of research software science (RSS) to elevate the RSE practice and its impact due to the complexity of research software. 
Overall, the relationship between HPC and software engineering continues to evolve at a great pace.
It is important to understand cost and quality trade-offs as new technologies become available as well as the unique aspects of HPC software development.

\section{Conclusions}
\label{sec:Conclusions}

The evolution of scientific software is a requirement to continue fulfilling sustainability aspects of science.
We have shown how the strategic expansion of certain modern software engineering practices can impact the overall development process of the QMCPACK framework targeting HPC systems.
We provide empirical data as evidence for how expanding CI efforts with a combination of cloud-based CI on free GitHub Actions runners and project self-maintained hardware resources for testing on NVIDIA and AMD GPUs has enabled QMCPACK readiness as exascale systems continue to be deployed.
In addition, the targeted quality strategy adds extra checks and refactoring efforts to sanitize critical code paths, track code testing coverage, and allow for multiple supported build configurations.
We outlined some of the important software engineering efforts in the QMCPACK software evolution roadmap: the removal of legacy code implementations, adding more unit tests to track coverage, modernizing current checkpoint-restart and I/O capabilities, refactoring compile-time configuration into runtime variants.
Our goal is to showcase how the targeted, predictive, proactive (rather than reactive), and selective application of modern software engineering practices have benefited a well-established HPC scientific code and community like QMCPACK.
We continue to advocate in the research software engineering (RSE) community that understanding the ``critical path" for a scientific software project is the first step in the successful and pragmatic integration of software engineering and scientific computing.
Publishing these types of experiences enriches the RSE community as a whole by showcasing the value and impact of their work.

\section*{Acknowledgment}
This research was supported by the Exascale Computing Project (17-SC-20-SC), a collaborative effort of the US Department of Energy Office of Science and the National Nuclear Security Administration.

\bibliographystyle{IEEEtran}
\bibliography{IEEEabrv,references}

\appendices

\section{Artifact Description}
\label{ap1:Artifact}

The artifacts presented in this paper are hosted or can be generated from the QMCPACK GitHub public repository ~\url{https://github.com/QMCPACK/qmcpack}. 


Docker container images for CI
\begin{itemize}
    \item Ubuntu 20: \url{https://hub.docker.com/repository/docker/williamfgc/qmcpack-ci/general}
    \item Ubuntu 22: \url{https://hub.docker.com/r/walshmm/qmcpack-ci/tags}
    \item Spack CentOS: 
    \url{https://hub.docker.com/r/walshmm/qmcpack-ci/tags}
\end{itemize}

Codecov report for QMCPACK test coverage \url{https://app.codecov.io/gh/QMCPACK/qmcpack}, 52.9\% or 49,351 out of 94,746 lines covered as of April 2023.

Scripts used to generate CI logs using GitHub CLI \url{https://code.ornl.gov/wfg/reproducibility-scripts/-/tree/main/QMCPACK/GitHubCLI.jl}

Tracking issues on QMCPACK repository:
\begin{itemize}
    \item Adoption of GitHub Actions CI: \url{https://github.com/QMCPACK/qmcpack/issues/3020}
    \item Addressing memory leaks:
    \url{https://github.com/QMCPACK/qmcpack/issues/3312}
    \item Removal of legacy CUDA driver implementation:
    \url{https://github.com/QMCPACK/qmcpack/issues/3856}
\end{itemize}


\end{document}